\documentclass [10pt]{article}
\usepackage{amsmath}
\usepackage{amsfonts}
\usepackage{amssymb}
\usepackage{graphicx}
\makeatletter \oddsidemargin  0in \evensidemargin 0in
\newcommand{\singlespacing}{\let\CS=\@currsize\renewcommand{\baselinestretch}{1.5}\tiny\CS}
\makeatletter \oddsidemargin  -.1in \evensidemargin -.1in
\newcommand{\doublespacing}{\let\CS=\@currsize\renewcommand{\baselinestretch}{1.35}\tiny\CS}

\def\@citex[#1]#2{\if@filesw\immediate\write\@auxout{\string\citation{#2}}\fi
  \def\@citea{}\@cite{\@for\@citeb:=#2\do
    {\@citea\def\@citea{,\linebreak[0]\hskip0pt plus .2em}%
      \@ifundefined{b@\@citeb}%
    {{\bf ?}\@warning{Citation `\@citeb' on page \thepage\space undefined}}%
      \hbox{\csname b@\@citeb\endcsname}}}{#1}}

\newtheorem{rule-def}[theorem]{Rule}

\begin{document}

\title{\bf Study of controlled dense coding with some discrete tripartite and quadripartite states}
\author{ Sovik Roy$^{1,2}$\thanks{Corresponding Author:\,sovik1891@gmail.com}, Biplab Ghosh$^{3}$\thanks{quantumroshni@gmail.com}\\
$^1$ Department of Mathematics, Techno India,Salt Lake, Sector V, Kolkata-700091,India\\
$^2$ S. N.Bose National Centre for Basic Sciences, Salt Lake, Kolkata-700098, India
\\
$^3$ Department of Physics, Vivekananda College for Women, Barisha, Kolkata-700008, India\\
}
\maketitle{}

\begin{abstract}
The paper presents the detailed study of controlled dense coding scheme for different types of $3$ and $4-$ particle states. It consists of GHZ state, GHZ type states, Maximal Slice state, Four particle GHZ state and W class of states. It is shown that GHZ-type states can be used for controlled dense coding in a probabilistic sense. We have shown relations among parameter of GHZ type state, concurrence of the shared bipartite state by two parties with respect to GHZ type and Charlie's measurement angle $\theta$. The GHZ states as a special case of maximal sliced states, depending on parameters, have also been considered here. We have seen that tripartite W state and quadripartite W state cannot be used in controlled dense coding whereas $\vert W_{n}\rangle_{ABC}$ states can be used probabilistically. Finally, we have investigated controlled dense coding scheme for tripartite qutrit states. 
\end{abstract}
\section{Introduction:}
Quantum entanglement plays a crucial role in quantum information processing. It is efficiently used as a key resource in several communication protocols for sending quantum as well as classical information from a sender to a receiver. Dense coding (\cite{bennwies}, \cite{matwekwze}) is one such exhibitions of entanglement in quantum communication. Quantum dense coding (QDC) deals with efficient information transfer from a sender to a receiver utilizing an entangled channel between the two. Moreover, QDC is the process of transmitting two-bits of classical information by sending only one qubit if the sender and the receiver share a maximally entangled state. In this respect, Hao et.al gave one protocol of controlled dense coding \cite{haoliguo} where they considered GHZ - states \cite{GHZ} and showed that the three qubits of GHZ state was shared by three parties Alice (the sender) , Bob (the receiver) and third party Charlie is the controller of the scheme. In \cite{fuxializh}, Fu et. al studied controlled dense coding with four particle non maximally entangled state. There they showed that the transmission of bits from party-$2$ to party-$3$ is controlled by party-$1$ and party-$4$'s local measurements as well. Our goal is to realize the controlled dense coding scheme in a tripartite  maximally entangled state in higher dimension. With this motivation, in this paper, we mainly emphasize on controlled dense coding with GHZ states, since these states can be generated in the laboratory and have been demonstrated experimentally using two pairs of entangled photons \cite{bopadaweze} and also one can make optimal distillation of these states \cite{acin}. We therefore we have used different forms of GHZ -class states, like GHZ -type, quadripartite GHZ states and GHZ states as a special case of maximal slice states \cite{carsud}. In due process, we also have studied controlled dense coding on states of W-class \cite{liqiu}. W-class of states are important from various aspects. Three particle W state $\frac{\vert 100\rangle + \vert 010\rangle + \vert 001\rangle}{\sqrt{3}}$ can be generated using super conducting phase qubits \cite{neelay}.The three particle GHZ and W states are fully characterized using quantum state tomography \cite{Acin}. Also four particle W states have been considered here for the study of controlled dense coding. These quadripartite W states can also be generated by parametric down conversion method \cite{mikami}. Study of controlled dense coding is then made upon tripartite qutrit state of GHZ form in this paper. Such qutrit states are generated via adiabatic passage of dark states \cite{biaobiao}. The paper is therefore organized as follows.\\In section $2.1$ we have given an overview of controlled dense coding scheme with various classifications of GHZ states. In section $2.2$ we made a study with GHZ type states. Since we know that GHZ type states are not useful for perfect dense coding and for these states probabilistic dense coding was studied by \cite{patiparaagra}, hence study of controlled dense coding was made with GHZ type states in a probabilistic manner. Section $2.3$ deals with Maximal slice states whereas Four particle GHZ states were studied with respect to controlled dense coding in section $2.4$. In sections $3.1$, $3.2$ and $3.3$ we used controlled dense coding with W-class of states. Finally in section $4$ the controlled dense coding has been done with tripartite qutrit states,  which in turn is followed by summary and discussion in section $5$.
\section{Controlled dense coding with GHZ - class states:}
\subsection{GHZ states:}
We first consider GHZ state of the form
\begin{equation}
\vert GHZ\rangle_{ABC} = \frac{1}{\sqrt{2}}\: (\:\vert 001 \rangle _{ABC} + \vert 110 \rangle_{ABC}).\: \label{ghz1}
\end{equation}
where the first qubit belongs to Alice (A), second belongs to Bob (B) and the third qubit is taken by Charlie (C). As described in \cite{haoliguo}, the scheme of `controlled dense coding', we assume that Charlie measures his qubit under the basis
\begin{eqnarray}
\vert +\rangle _{C} = \cos \theta\:\vert 0 \rangle_{C}\: + \sin \theta \: \vert1 \rangle_{C},\nonumber\\
\vert - \rangle_{C} = \sin \theta\:\vert 0 \rangle_{C}\: - \cos \theta \: \vert 1 \rangle_{C}\ \label{charliebasis1}
\end{eqnarray}
where $\vert \: \sin \theta\:\vert \leq \vert \: \cos \theta \:\vert$. Then using (\ref{charliebasis1}), (\ref{ghz1}) can be expressed as
\begin{eqnarray}
\vert GHZ\rangle_{ABC} = \frac{1}{\sqrt{2}}\: (\vert +\rangle_{C}\:\vert \psi\rangle _{AB} + \vert -\rangle_{C} \vert \varphi\rangle_{AB})\: \label{ghz1a}
\end{eqnarray}
where
\begin{eqnarray}
\vert \psi\rangle _{AB} = \sin \theta\:\vert 00 \rangle_{AB}\: + \cos \theta \: \vert 11 \rangle_{AB},\nonumber\\
\vert \varphi\rangle_{AB} = \sin \theta\:\vert 11 \rangle_{AB}\: - \cos \theta \: \vert 00 \rangle_{AB}.\ \label{nmestates1}
\end{eqnarray}
von-Neumann measurement of qubit C gives either $\vert +\rangle_{C}$ or $\vert -\rangle_{C}$, (each occurring with equal probability $\frac{1}{2}$), depending upon which, Alice and Bob shares either of the above two bipartite non - maximally entangled states from (\ref{nmestates1}) respectively. If for instance, Charlie gets his read-out as $\vert +\rangle_{C}$, he informs Alice about his outcome so that Alice knows she shares non-maximally entangled state $\vert \psi\rangle _{AB}$ with Bob now (whereas Bob does not know this). Alice introduces an auxilliary qubit $\vert 0\rangle_{aux}$ and performs an unitary operation on her qubit A and the auxilliary qubit (with respect to the collective operation under the basis ($\vert 00\rangle_{Aaux}, \vert 10\rangle_{Aaux}, \vert 01\rangle_{Aaux}, \vert 11\rangle_{Aaux}$). The collective unitary operation $U_{1}\otimes I_{B}$ transforms the state $\vert \psi\rangle _{AB}\otimes \vert 0\rangle_{aux}$ to
\begin{eqnarray}
\vert \psi\rangle_{ABaux} = \sqrt{2}\:\cos\:\theta\:\lbrace\: \frac{1}{\sqrt{2}}\:(\vert 00 \rangle_{AB}\:+\:\vert 11 \rangle_{AB}\:)\:\rbrace\:\ \otimes \vert 0 \rangle_{aux} \nonumber{}\\ 
+ \sin\:\theta\:\sqrt{\:1-\frac{\cos^{2}\:\theta}{\sin^{2}\:\theta}}\:\vert 10 \rangle_{AB}\:\otimes\:\vert 1 \rangle_{aux}. \label{transformed1}
\end{eqnarray}
where,
\begin{eqnarray}
U_{1} = \left(%
\begin{array}{cccc}
\frac{\cos\:\theta}{\sin\:\theta}& 0& \sqrt{1-\frac{\cos^{2}\:\theta}{\sin^{2}\:\theta}}& 0\\
0& 1& 0& 0\\
0& 0& 0& -1\\
\sqrt{1-\frac{\cos^{2}\:\theta}{\sin^{2}\:\theta}}& 0& -\frac{\cos\:\theta}{\sin\:\theta}& 0
\end{array}%
\right).\label{unitarymatrix1}
\end{eqnarray}
When Alice performs one of the four operations $\lbrace\:I, \sigma^{x}, i\sigma^{y}, \sigma^{z}\:\rbrace$ (where $I$ is the idenstity operator and $\sigma^{m}$'s $m = x, y, z$ are three Pauli spin operators) on her qubit and sends the qubit to Bob where Bob in turn carries out a controlled-NOT operation on his qubit, then it has been shown in \cite{haoliguo} that average number of bits transmitted from Alice to Bob at the cost of one GHZ state of the form (\ref{ghz1}) is  
\begin{equation}
Av_{transm} = 1 + P_{success} \times 1 = 1 + 2\: \vert \cos\:\theta\vert^{2} \label{success1}
\end{equation}
provided the von-Neumann measurement performed by Alice on the auxilliary qubit results in the outcome $\vert 0\rangle_{aux}$, as in that case Alice and Bob will share maximally entangled state $\frac{1}{\sqrt{2}}[\vert 00\rangle_{AB} + \vert 11\rangle_{AB}]$, which is clear from (\ref{transformed1}). But when Alice's von-Neumann readout is $\vert 1\rangle_{aux}$ then only one bit is transferred. The maximal value of $\vert\cos\:\theta\vert$ is $\frac{1}{\sqrt{2}}$, which corresponds to the maximally entangled state $\frac{1}{\sqrt{2}}\:(\vert 00 \rangle_{12}\:+\vert 11 \rangle_{AB}\:)$ and two bits are transmitted  by one qubit, which depends on the adjustment of the value of $\theta$ by Charlie (the controller of the scheme). If however Charlie gets his measurement result as $\vert -\rangle_{C}$, the shared state between Alice and Bob is then $\vert \varphi\rangle_{AB}$ from (\ref{nmestates1}). Proceeding in a similar way we can show that the average number of bit transmitted is then $1 + 2\: \vert \sin\:\theta\vert^{2}$.\\
Now we choose another unitary matrix $U_{2}$ which is different from that of $U_{1}$ of (\ref{unitarymatrix1}), as 
\begin{eqnarray}
U_{2} = \left(%
\begin{array}{cccc}
1& 0& 0& 0\\
0& \frac{\cos\:\theta}{\sin\:\theta}& \sqrt{1-\frac{\cos^{2}\:\theta}{\sin^{2}\:\theta}}& 0\\
0& -\sqrt{1-\frac{\cos^{2}\:\theta}{\sin^{2}\:\theta}}& \frac{\cos\:\theta}{\sin\:\theta}& 0\\
0& 0& 0& 1
\end{array}%
\right).\label{unitarymatrix2}
\end{eqnarray}
Then with same study as above we see that depending upon Charlie's measurement $\vert +\rangle_{C}$ , introduction of Alice's auxilliary qubit $\vert 1\rangle_{aux}$  and using collective unitary operation $U_{2} \otimes \vert 1\rangle_{aux}$, the average number of bit transmitted is now $1 + 2\: \vert \cos\:\theta\vert^{2}$ at the cost of one GHZ state of the form (\ref{ghz1}).
\\
The study of controlled dense coding \cite{haoliguo} is now performed on various classifications of GHZ states as shown below 
\begin{eqnarray}
G_{1}=\frac{1}{\sqrt{2}}(\vert 010 \rangle _{ABC} + \vert 101 \rangle_{ABC}),\nonumber{}\\
G_{2}=\frac{1}{\sqrt{2}} (\vert 010 \rangle _{ABC} - \vert 101 \rangle_{ABC}),\nonumber{}\\
G_{3}=\frac{1}{\sqrt{2}}(\vert 001 \rangle _{ABC} - \vert 110 \rangle_{ABC}),\nonumber{}\\
G_{4}=\frac{1}{\sqrt{2}}(\vert 001 \rangle _{ABC} + \vert 110 \rangle_{ABC}),\nonumber{}\\
G_{5}=\frac{1}{\sqrt{2}}(\vert 100 \rangle _{ABC} - \vert 011 \rangle_{ABC}),\nonumber{}\\
G_{6}=\frac{1}{\sqrt{2}}(\vert 100 \rangle _{ABC} + \vert 011 \rangle_{ABC}),\nonumber{}\\
G_{7}=\frac{1}{\sqrt{2}}(\vert 000 \rangle _{ABC} - \vert 111 \rangle_{ABC})\label{ghz1b}
\end{eqnarray} 
with proper choice of unitary matrix. These states of (\ref{ghz1b}) can be obtained when Charlie renames his basis applying Pauli $\sigma^{x}$, say, which is a local unitary.\\
We put our study results of controlled dense coding with states of (\ref{ghz1b}) in the table below showing the average number of bits transmitted for each of the above states with respect to von-Neumann measurements of Charlie.
\begin{center}
\begin{tabular}{|c|c|c|}
\hline States&  Charlie's von Neumann readout&  $Av_{transm}$ \\
\hline $G_{1}$ &  $\vert + \rangle_{C}$ &  $1 + 2
\vert \sin\:\theta\vert^{2}$ \\
\hline $G_{2}$ &  $\vert - \rangle_{C}$ &   $1 + 2\: \vert \cos\:\theta\vert^{2}$\\
\hline $G_{3}$ &  $\vert - \rangle_{C}$ &   $1 + 2\: \vert \sin\:\theta\vert^{2}$\\
\hline $G_{4}$ &  $\vert + \rangle_{C}$ &  $1 + 2\: \vert \cos\:\theta\vert^{2}$\\
\hline $G_{5}$ &  $\vert - \rangle_{C}$ &   $1 + 2\: \vert \sin\:\theta\vert^{2}$\\
\hline $G_{6}$ &  $\vert + \rangle_{C}$ &   $1 + 2\:\vert \cos\:\theta\vert^{2}$\\
\hline $G_{7}$ &  $\vert - \rangle_{C}$ &  $1 + 2\:\vert \cos\:\theta\vert^{2}$\\
\hline
\end{tabular}
\end{center}\vskip0.5cm
\noindent We see from above table that for some states like $G_{1}, G_{4}, G_{6}$ the number of bits transmitted is $1 + 2\: \vert \sin\:\theta\vert^{2}$ and for the other states $G_{2}, G_{3}, G_{5}, G_{7}$, the number of bits transmitted is $1 + 2\: \vert \cos\:\theta\vert^{2}$. The results are then shown graphically below\\ 
\begin{figure}[!ht]
\centering
\resizebox{8cm}{8cm}{\includegraphics{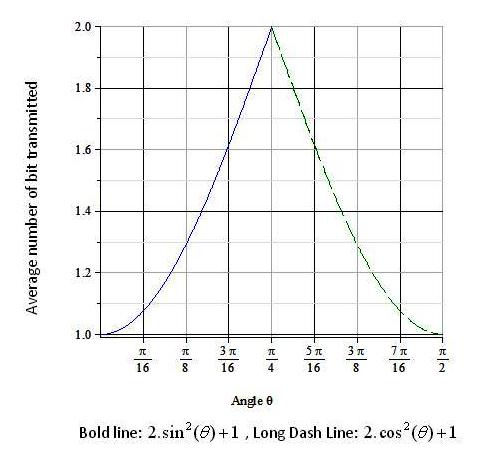}}
\caption{\footnotesize The figure represents average number of bits transmitted against the angle $\theta$. The solid line represents $2\sin^{2}\theta + 1$ bits are transmitted for some GHZ forms for $0 \leqslant \theta \leqslant \frac{\pi}{4}$. The dashed line represents $2\cos^{2}\theta + 1$ bits are transmitted for some other GHZ forms for $\frac{\pi}{4} \leqslant \theta \leqslant \frac{\pi}{2}$ whereas optimum value of bit transmission is $2$.}
\end{figure}
\subsection{GHZ-type states:}
It is a well known fact that the states like $\frac{1}{\sqrt{3}}\: (\:\sqrt{2}\:\vert 000 \rangle _{ABC} + \vert 111 \rangle_{ABC})\:$ cannot be used for perfect super dense coding \cite{liqiu}. Pati et. al showed in \cite{patiparaagra} that with such states dense coding is performed probabilistically. We  take general form of GHZ type state as given in \cite{patiparaagra}
\begin{equation}
\vert \zeta \rangle_{ABC} = L\: (\:\vert 000 \rangle _{ABC} + l\:\vert 111 \rangle_{ABC})\: \label{pati1}
\end{equation}
where $L=\frac{1}{\sqrt{1+l^{2}}}$ and $l>0$ (considering $l$ to be real). Charile here chooses his measurement basis from (\ref{charliebasis1}). With respect to these bases, (\ref{pati1}) can be expressed as
\begin{equation}
\vert \zeta \rangle_{ABC} = \frac{1}{\sqrt{1+l^{2}}}\:[(\vert s_{1}\rangle_{AB} \vert +\rangle_{C}\:
+\vert s_{2}\rangle_{AB}\:\vert -\rangle_{C}]\label{pati2}
\end{equation} 
where
\begin{eqnarray}
\vert s_{1}\rangle_{AB}=\cos\:\theta \vert 00\rangle_{AB}+l\:\sin\theta \vert 11\rangle_{AB},\nonumber{}\\
\vert s_{2}\rangle_{AB}=\sin\theta \vert 00\rangle_{AB}-l\cos\theta\vert 11\rangle_{AB}.\label{pati3}
\end{eqnarray}
If von-Neumann measurement of Charlie gives $\vert +\rangle_{C}$, the non maximally entangled state shared between Alice and Bob is $\vert s_{1}\rangle_{AB}$. After Alice introduces auxilliary qubit $\vert 0\rangle_{aux}$ and considers the unitary operator \\
\begin{eqnarray}
U^{/}_{1} = \left(%
\begin{array}{cccc}
\frac{\sin\:\theta}{\cos\:\theta}& 0& \sqrt{1-\frac{\sin^{2}\:\theta}{\cos^{2}\:\theta}}& 0\\
0& 1& 0& 0\\
0& 0& 0& -1\\
\sqrt{1-\frac{\sin^{2}\:\theta}{\cos^{2}\:\theta}}& 0& -\frac{\sin\:\theta}{\cos\:\theta}& 0
\end{array}%
\right)\label{unitarymatrix3}.
\end{eqnarray}\\
The collective unitary operation $U^{/}_{1}\otimes I_{B}$ transforms the state $\vert s_{1}\rangle_{AB} \otimes \vert 0\rangle_{aux}$ to\\
\begin{eqnarray}
\vert s_{1}\rangle_{ABaux} = \frac{\sin\:\theta}{L}{\lbrace L(\vert 00\rangle_{AB}+l\:\vert 11\rangle_{AB})\rbrace}\otimes\vert 0\rangle_{aux} \nonumber{}\\ 
+ \sqrt{1-\frac{\sin^{2}\theta}{\cos^{2}\theta}}\:\cos\:\theta \vert 10\rangle_{AB}\otimes\vert 1\rangle_{aux}. \label{transformed2}
\end{eqnarray}\\
Alice's von-Neumann measurement of $\vert 1\rangle_{aux}$ shows that only one bit is transferred from Alice to Bob whereas the measurement result of $\vert 0\rangle_{aux}$ shows that Alice and Bob shares $L\: (\:\vert 00 \rangle _{AB} + l\:\vert 11 \rangle_{AB})$. The average number of bit transmitted is then $1 + \frac{\vert \sin\:\theta \vert^{2}}{L^{2}}$, and the scheme is successful only if $\theta\:=\sin^{-1}\:(\pm\frac{1}{\sqrt{1+l^{2}}})$. As described in \cite{patiparaagra}, Alice applies any one of the four unitary operators $\lbrace I, \sigma^{x}, i\sigma^{y},\sigma^{z} \rbrace$. The shared state between Alice and Bob then takes the form
\begin{eqnarray}
L(\vert 00\rangle_{AB}+l\:\vert 11\rangle_{AB}),\nonumber{}\\ 
L(\vert 10\rangle_{AB}+l\:\vert 01\rangle_{AB}),\nonumber{}\\ 
L(-\vert 10\rangle_{AB}+l\:\vert 01\rangle_{AB}),\nonumber{}\\ 
L(\vert 00\rangle_{AB}-l\:\vert 11\rangle_{AB}).\nonumber{}\\ 
 \label{basis}
\end{eqnarray}
Alice sends her qubit to Bob. Bob performs a projection on to the basis spanned by the basis states $\lbrace \vert 00 \rangle , \vert 11\rangle \rbrace$ and $\lbrace \vert 01 \rangle , \vert 10\rangle \rbrace$. It is also shown in \cite{patiparaagra} that, Bob can extract two bits of classical information with a success probability is $\frac{2\:l^{2}}{1+l^{2}}$. Thus for maximally entangled state, $l=1$ and hence the success probability is unity. It is also clear that for maximally entangled state, Charlie's measurement angle is therefore $\pm \frac{\pi}{4}$. The above analysis is shown in the table below
\begin{center}
\begin{tabular}{|c|c|c|c|}
\hline $l$&  $\theta$&  shared state&  Success probability = $\frac{2\:l^{2}}{1+l^{2}}$\\
\hline $0$ &  $\pm \frac{\pi}{2}$ &  $\vert 00 \rangle_{AB}$& $0$ \\
\hline $1$ &  $\pm \frac{\pi}{4}$ &   $\frac{1}{2}[\vert 00\rangle_{AB} + \vert 11\rangle_{AB}]$& $1$\\
\hline
\end{tabular}
\end{center}\vskip0.5cm
\noindent We see that with states of the form (\ref{pati1}), controlled dense coding is thus achieved in a probabilistic way. In this case we also find a relation between the parameter $l$ and angle $\theta$ as
\begin{equation}
\theta\:=\tan^{-1}\frac{1}{l}.\: \label{rel1}
\end{equation}
The relation (\ref{rel1}) is shown graphically below
\begin{figure}[!ht]
\centering
\resizebox{7cm}{7cm}{\includegraphics{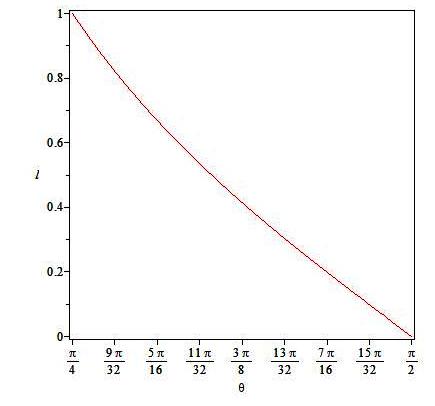}}
\caption{\footnotesize The figure shows that for $\frac{\pi}{4}\:\leq\:\theta\:\leq\:\frac{\pi}{2}$, we have $1\:\leq l\:\leq\: 0$.}
\end{figure}\\
From fig-2 it is clear that when $\theta=\frac{\pi}{4}$, $l=1$, the scheme is achieved successfully with success probability $1$.\\
We shall now study the entanglement property of the shared state $L\: (\:\vert 00 \rangle _{AB} + l\:\vert 11 \rangle_{AB})$ between Alice and Bob and how it is dependent on the parameter $l$. The concurrence for a bipartite state $\rho_{AB}$ is defined as \cite{wootters}
\begin{equation}
C\:=\: max\:\lbrace 0, \lambda_{1}-\lambda_{2}-\lambda_{3}-\lambda_{4}\:
\rbrace \label{concur1}
\end{equation}
where $\lambda's$ are the square root of eigenvalues of $\rho\tilde{\rho}$ in decreasing order. The spin - flipped density matrix $\tilde{\rho}$ is defined as
\begin{equation}
\tilde{\rho}\:=(\sigma^{y}_{A}\otimes\sigma^{y}_{B})\rho^{*}(\sigma^{y}_{A}\otimes\sigma^{y}_{B})
\rbrace \label{spinflip}
\end{equation}
Using (\ref{concur1}) and (\ref{spinflip}) we calculate the concurrence $C$ of the shared state  $L\: (\:\vert 00 \rangle _{AB} + l\:\vert 11 \rangle_{AB})$ which is 
\begin{equation}
C = \vert 2\:L^{2}\:l \vert \label{rel2}
\end{equation}
Then from (\ref{rel1}) and (\ref{rel2}) we get
\begin{equation}
C\:=\vert \sin\:2\:\theta \vert
\label{concur2}
\end{equation}
The plot of C versus $\theta$ gives
\begin{figure}[!ht]
\centering
\resizebox{7cm}{7cm}{\includegraphics{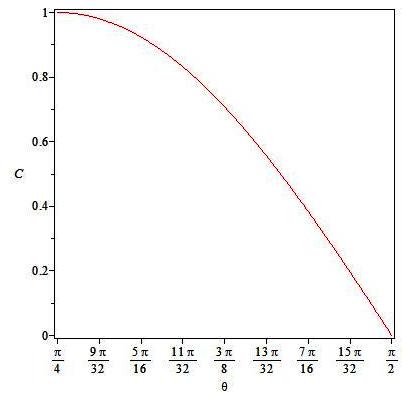}}
\caption{\footnotesize The figure shows that for $\frac{\pi}{4}\:\leq\:\theta\:\leq\:\frac{\pi}{2}$, we have $1\:\leq C\:\leq\: 0$.}
\end{figure}\\
The above analysis shows that, before dense coding is executed between Alice and Bob, what state they will share, is also controlled by Charlie. The protocol of dense coding is done successfully provided Charlie fixes his measurement angle to $\theta\:=\frac{\pi}{4}$, as in this case only Alice and Bob shares maximally entangled state and the success probability is therefore $1$.
\subsection{GHZ state as a special case of MS state:}
The maximal slice (MS) state is defined in \cite{carsud} as
\begin{equation}
\vert \phi\rangle^{MS}_{ABC}\:= \frac{1}{\sqrt{2}}[\vert 000\rangle_{ABC}+\:\cos\:\delta\vert 110 \rangle_{ABC}+\:\sin\:\delta\vert 111 \rangle_{ABC}]
\label{msstate1}
\end{equation}
As before if Charlie chooses his basis from (\ref{charliebasis1}), the state shared between Alice and Bob is either of the following two,  depending upon Charlie's measurement basis $\lbrace \vert +\rangle_{C}, \vert -\rangle_{C}\rbrace$.
\begin{eqnarray}
\vert \phi\rangle^{MS1}_{AB}\:= \sin\:\theta\:\vert 00\rangle_{AB}-\:\cos\:\theta\:\sin\:\delta\vert 11 \rangle_{AB}+\:\sin\:\theta\:\cos\:\delta\vert 11 \rangle_{AB} \nonumber{}\\
\vert \phi\rangle^{MS2}_{AB}\:= \cos\:\theta\:\vert 00\rangle_{AB}+\:\cos\:\theta\:\cos\:\delta\vert 11 \rangle_{AB}+\:\sin\:\theta\:\sin\:\delta\vert 11 \rangle_{AB} \nonumber{}\\
\label{msstate2}
\end{eqnarray}
If Alice and Bob shares the state $\vert \phi\rangle^{MS1}_{AB}$ then after Alice introduces auxilliary qubit $\vert 0\rangle_{aux}$, choosing unitary operator (\ref{unitarymatrix3}), the collective unitary operation $U^{/}_{1}\otimes I_{B}$  transforms the state $\vert \phi\rangle^{MS1}_{AB} \otimes \vert 0 \rangle_{aux}$ to
\begin{eqnarray}
\vert \phi\rangle^{MS1}_{ABaux}\:=\sin\:\theta\:[\vert 00\rangle_{AB}+\frac{\cos\:\theta}{\sin\:\theta}\:\cos\:\delta\:\vert 11\rangle_{AB}+\sin\:\delta\vert 11\rangle_{AB}]\otimes \vert 0\rangle_{aux}\nonumber{}\\
+ \cos\:\theta\:\sqrt{1-\frac{\sin^{2}\theta}{\cos^{2}\theta}}\vert 10\rangle_{AB}\otimes \vert 1\rangle_{aux}
\label{msstate1trans}
\end{eqnarray}
Alice's von-Neumann measurement result of $\vert 0\rangle_{aux}$ gives that the non-maximally entangled shared state between Alice and Bob is~~ $\sin\:\theta\:[\vert 00\rangle_{AB}+\frac{\cos\:\theta}{\sin\:\theta}\:\cos\:\delta\:\vert 11\rangle_{AB}+\sin\:\delta\vert 11\rangle_{AB}]$. If Charlie fixes his measurement angle to $\theta = \frac{\pi}{4}$, the shared state takes the form $\frac{1}{\sqrt{2}}[\vert 00\rangle_{AB}+(\cos\:\delta +\sin\:\delta)\vert 11\rangle_{AB}]$. The normalization condition then gives $\delta$ should be equal to $\frac{n\:\pi}{2}$.  With the fixed values of $\theta$ and $\delta$, the MS state (\ref{msstate1}) takes the GHZ form i.e. $\frac{1}{\sqrt{2}}[\vert 000\rangle_{ABC}+\vert 111\rangle_{ABC}]$, the controlled dense coding with which was already shown by \cite{haoliguo}.
\subsection{Four particle GHZ states:}
In \cite{fuxializh} Fu et.al have shown the protocol of controlled dense coding with a non - maximally entangled state. We in this section use the Fu protocol for four particle GHZ state given by
\begin{equation}
\vert GHZ\rangle_{PABC}\:= \frac{1}{\sqrt{2}}[\vert 0000\rangle_{PABC}+ \vert 1111 \rangle_{PABC}].
\label{ghz4}
\end{equation}
where P, A, B, C respectively represents Paul, Alice, Bob and Charlie. Now Charlie chooses his measurement basis as given in (\ref{charliebasis1}). The state (\ref{ghz4}) is therefore expressed as
\begin{equation}
\vert GHZ\rangle_{PABC}\:= \frac{1}{\sqrt{2}}[\:\vert \varsigma\rangle_{PAB}\vert +\rangle_{C}+ \vert \tau \rangle_{PAB}\vert -\rangle_{C}\:]
\label{ghz4a}
\end{equation}
such that
\begin{eqnarray}
\vert \varsigma\rangle_{PAB}\:= \:\cos\:\theta \vert 000\rangle_{PAB} + \:\sin\:\theta \vert 111 \rangle_{PAB},\nonumber{}\\
\vert \tau\rangle_{PAB}\:= \:\sin\:\theta \vert 000\rangle_{PAB} - \:\cos\:\theta \vert 111 \rangle_{PAB}.\nonumber{}\\
\label{ghz4a}
\end{eqnarray}
Paul now chooses a new basis 
\begin{eqnarray}
\vert +\rangle _{P} = \cos \varepsilon\:\vert 0 \rangle_{P}\: + \sin \varepsilon \: \vert1 \rangle_{P},\nonumber\\
\vert - \rangle_{P} = \sin \varepsilon\:\vert 0 \rangle_{P}\: - \cos \varepsilon \: \vert 1 \rangle_{P}.\ \label{paulbasis2}
\end{eqnarray}
and carries out a unitary operation on his qubit. Suppose Charlie's measurement result gives $\vert +\rangle_{C}$, then Paul, Alice and Bob shares the state $\vert \varsigma\rangle_{PAB}$. With respect to Paul's basis (\ref{paulbasis2}),  $\vert \varsigma\rangle_{PAB}$ is then expressed as, 
\begin{equation}
\vert \varsigma\rangle_{PAB}\:= \vert \mu \rangle_{AB}\otimes \vert +\rangle_{P} + \vert \nu \rangle_{AB}\otimes \vert -\rangle_{P}
\end{equation}\label{ghz4b}
where
\begin{eqnarray}
\vert \mu \rangle_{AB}=\cos\:\theta\:\cos\:\varepsilon \vert 00\rangle_{AB}+\sin\:\theta\:\sin\:\varepsilon \vert 11\rangle_{AB},\nonumber{}\\
\vert \nu \rangle_{AB}=\cos\:\theta\:\sin\:\varepsilon \vert 00\rangle_{AB}-\sin\:\theta\:\cos\:\varepsilon \vert 11\rangle_{AB}\nonumber{}.\\
\end{eqnarray}\label{ghz4b1}
For Paul's local measurement result $\vert +\rangle_{P}$, the shared state between Alice and Bob is $\vert \mu \rangle_{AB}$. Now Alice introduces one auxilliary qubit $\vert 0\rangle_{aux}$. We consider unitary operator
\begin{eqnarray}
U_{3} = \left(%
\begin{array}{cccc}
\frac{\sin\:\theta\:\sin\:\varepsilon}{\cos\:\theta\:\cos\:\varepsilon}& 0& \sqrt{1-\frac{\sin^{2}\:\theta\:\sin^{2}\:\varepsilon}{\cos^{2}\:\theta\:\cos^{2}\:\varepsilon}}& 0\\
0& 1& 0& 0\\
-\sqrt{1-\frac{\sin^{2}\:\theta\:\sin^{2}\:\varepsilon}{\cos^{2}\:\theta\:\cos^{2}\:\varepsilon}}& 1& \frac{\sin\:\theta\:\sin\:\varepsilon}{\cos\:\theta\:\cos\:\varepsilon}& 0\\
0& 0& 0& -1\\
\end{array}%
\right)\label{unitarymatrix4}
\end{eqnarray}
The collective unitary operation of $U_{3}\otimes I_{B}$ transforms $\vert \mu \rangle_{AB} \otimes \vert 0\rangle_{aux}$ to
\begin{eqnarray}
\vert \mu \rangle_{ABaux}\:=\:\sin\:\theta\:\sin\:\varepsilon\:[\vert 00\rangle_{AB}+\vert 11\rangle_{AB}]\otimes \vert 0\rangle_{aux}\nonumber{}\\
-\sqrt{1-\frac{\sin^{2}\:\theta\:\sin^{2}\:\varepsilon}{\cos^{2}\:\theta\:\cos^{2}\:\varepsilon}}\:\cos\:\theta\:\cos\:\varepsilon\:\vert 00\rangle_{AB}\otimes \vert 1\rangle_{aux}
\label{ghz4c}
\end{eqnarray}
The von-Neumann measurement outcome $\vert 0\rangle_{aux}$ of Alice shows that the non-maximally entangled state shared between Alice and Bob is therefore
\begin{equation}
\sin\:\theta\:\sin\:\varepsilon\:(\vert 00\rangle_{AB}+\vert 11\rangle_{AB}).
\label{ghz4d}
\end{equation}\\
Using \cite{wootters}, we calculate the concurrence $C_{1}$ of (\ref{ghz4d}), which is given by
\begin{equation}
C_{1}=\:2\:\sin^{2}\theta\:\sin^{2}\varepsilon.
\label{concur2}
\end{equation}
The above relation (\ref{concur2}) is shown graphically below
\begin{figure}[!ht]
\centering
\resizebox{7cm}{7cm}{\includegraphics{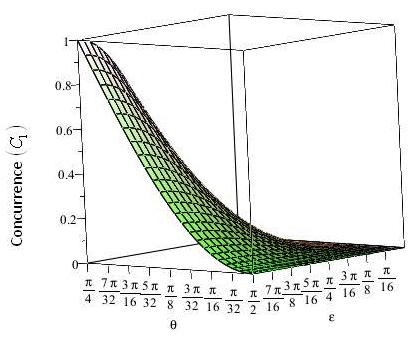}}
\caption{\footnotesize Here $0 \leq C_{1} \leq 1$ whereas $0 \leq \theta \leq \frac{\pi}{4}$ and $0 \leq \varepsilon \leq \frac{\pi}{2}$. It is clear that when $\theta = \frac{\pi}{4}$ and $\varepsilon = \frac{\pi}{2}$, concurrence is maximum i.e. $1$}
\end{figure}\\ \newpage
Hence when Paul fixes his measurement angle to $\varepsilon = \frac{\pi}{2}$, the state shared between Alice and Bob is ~~$\sin\:\theta\:[\vert 00\rangle_{AB} + \vert 11\rangle_{AB}]$ . When $\theta=\frac{\pi}{4}$, Alice and Bob shares maximally entangled state and two bits are transferred. Thus we see that, in this case what state will be shared by Alice and Bob is controlled both by Paul and Charlie as well as the bit transmission between Alice and Bob is controlled by them too.
\section{Controlled dense coding with W - class states:}
\subsection{W-state:}
Let us consider the state \cite{liqiu} in the form as
\begin{equation}
\vert W \rangle_{ABC}=\frac{1}{\sqrt{3}}[\vert 100\rangle_{ABC} + \vert 010\rangle_{ABC} + \vert 001\rangle_{ABC}]
\label{w1}
\end{equation}
When Charlie chooses his basis as given in (\ref{charliebasis1}), then the $W-$ state can be expressed as
\begin{eqnarray}
\vert W \rangle_{ABC}=\frac{1}{\sqrt{3}}[\:\vert +\rangle_{C}\vert \varpi\rangle_{AB}
+\:\vert -\rangle_{C}\vert \varpi^{/}\rangle_{AB}]
\label{w2}
\end{eqnarray}
where
\begin{eqnarray}
\vert \varpi\rangle_{AB}=\cos\:\theta \vert 10\rangle_{AB}+\cos\:\theta \vert 01\rangle_{AB}+\sin\:\theta \vert 00\rangle_{AB}\nonumber{}\\
\vert \varpi^{/}\rangle_{AB}=sin\:\theta \vert 10\rangle_{AB}+\sin\:\theta \vert 01\rangle_{AB}-\cos\:\theta \vert 00\rangle_{AB}\nonumber{}\\
\label{w3}
\end{eqnarray}
We assume that Charlie's von-Neumann measurement outcome is $\vert +\rangle_{C}$ so that Alice and Bob shares non maximally entangled state is $\vert \varpi\rangle_{AB}$. Alice introduces auxilliary qubit $\vert 0\rangle_{aux}$ and she considers unitary operator (\ref{unitarymatrix3}), such that the collective unitary operation $U^{/}_{1} \otimes I_{B}$ transforms the state $\vert \varpi\rangle_{AB} \otimes \vert 0\rangle_{aux}$ to
\begin{eqnarray}
\vert \varpi\rangle_{ABaux}=\:[\sin\:\theta\:\vert 01\rangle_{AB}+\frac{\sin^{2}\theta}{\cos\:\theta}\:\vert 00\rangle_{AB}+\cos\:\theta\:\vert 10\rangle_{AB}]\otimes \vert 0\rangle_{aux}\nonumber{}\\
+ [\sqrt{1-\frac{\sin^{2}\theta}{\cos^{2}\:\theta}}\cos\:\theta \vert 11\rangle_{AB} + \sqrt{1-\frac{\sin^{2}\theta}{\cos^{2}\:\theta}}\sin\:\theta \vert 10\rangle_{AB}]\otimes \vert 1\rangle_{aux}
\label{w4}
\end{eqnarray}\\
If now Alice gets her von-Neumann outcome as $\vert 0\rangle_{aux}$, this implies that Alice and Bob shares $[\:\sin\:\theta\:\vert 01\rangle_{AB}+\frac{\sin^{2}\theta}{\cos\:\theta}\:\vert 00\rangle_{AB}+\cos\:\theta\:\vert 10\rangle_{AB}\:]$.
Using (\ref{concur1}) we measure the concurrence ($C_{2}$) of the shared state and consequently find that\\
\begin{equation}
C_{2}= \sqrt{2}\:\vert\cos\:\theta\:\sin\:\theta\vert
\label{concur3}
\end{equation}
Then in the following we plot $C_{2}$ versus $\theta$.
\begin{figure}[!ht]
\centering
\resizebox{7cm}{7cm}{\includegraphics{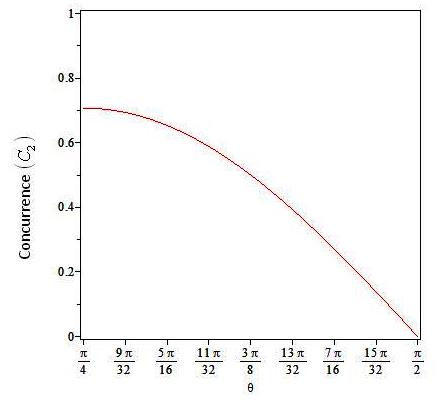}}
\caption{\footnotesize Here $\frac{\pi}{4} \leq \theta \leq \frac{\pi}{2}$ and $0.7 \leq C_{2} \leq 0$.}
\end{figure}\\
It is clear from the above figure that for $\theta$ assuming values from $\frac{\pi}{4}$ to $\frac{\pi}{2}$, concurrence of the shared state $[\:\sin\:\theta\:\vert 01\rangle_{AB}+\frac{\sin^{2}\theta}{\cos\:\theta}\:\vert 00\rangle_{AB}+\cos\:\theta\:\vert 10\rangle_{AB}\:]$ never reaches its maximum value i.e. $1$, rather it lies in the range $[0.7, 0]$. In other words Alice and Bob never share maximally entangled state for any value of $\theta$, which is the primary requirement for controlled dense coding. Hence the state (\ref{w1}) is not suitable for controlled dense coding.
\subsection{Four particle W state:}
Consider now a state of the form as \cite{liqiu}
\begin{equation}
\vert W \rangle_{PABC}=\frac{1}{\sqrt{4}}[\vert 1000\rangle_{PABC} + \vert 0100\rangle_{PABC} + \vert 0010\rangle_{PABC} + \vert 0001\rangle_{PABC}]
\label{W1}
\end{equation}
where P, A, B and C respectively are Paul, Alice, Bob, Charlie as before. We use now scheme of \cite{fuxializh} again.  The state (\ref{W1}) is then expressed in terms of Charlie's measurement basis $\lbrace\:\vert +\rangle_{C},\vert -\rangle_{C}\:\rbrace$ from (\ref{charliebasis1}) and consequently we get
\begin{equation}
\vert W \rangle_{PABC}=\frac{1}{\sqrt{2}}[\vert t_{1}\rangle_{PAB}\:\vert +\rangle_{C} +\vert t_{2}\rangle_{PAB}\: \vert -\rangle_{C})].
\label{W2}
\end{equation}
where
\begin{eqnarray}
\vert t_{1}\rangle_{PAB}=\frac{\cos\theta(\vert 100\rangle_{PAB}+\vert 010\rangle_{PAB}+\vert 001\rangle_{PAB})+\sin\theta\:\vert 000\rangle_{PAB}}{\sqrt{2}},\nonumber{}\\
\vert t_{2}\rangle_{PAB}=\frac{\sin\theta(\vert 100\rangle_{PAB}+\vert 010\rangle_{PAB}+\vert 001\rangle_{PAB})-\cos\theta\:\vert 000\rangle_{PAB}}{\sqrt{2}}.\label{W2a}
\end{eqnarray}
Again Paul chooses now his basis $\lbrace\:\vert +\rangle_{P},\vert -\rangle_{P}\:\rbrace$ from (\ref{paulbasis2}) and expresses (\ref{W2}) in terms of his basis elements as 
\begin{eqnarray}
\vert W \rangle_{PAB}=\frac{1}{\sqrt{2}}[\vert +\rangle_{P}\:\vert t^{/}_{1}\rangle_{AB}
+\vert -\rangle_{P}\vert t^{/}_{2}\rangle_{AB}]
\end{eqnarray}\label{W2b}
where
\begin{eqnarray}
\vert t^{/}_{1}\rangle_{AB} = \sin\:(\theta+\varepsilon) \vert 00\rangle_{AB} + \cos\:\theta\:\cos\:\varepsilon\:(\vert 10\rangle_{AB} + \vert 01\rangle_{AB}),\nonumber{}\\
\vert t^{/}_{2}\rangle_{AB}=-\cos\:(\theta+\varepsilon) \vert 00\rangle_{AB} + \cos\:\theta\:\sin\:\varepsilon\:(\vert 10\rangle_{AB} + \vert 01\
\rangle_{AB}).\nonumber{}\\
\end{eqnarray}\label{W2b}
If suppose we assume that Paul's von-Neumann measurement outcome results in $\vert +\rangle_{P}$, then the shared state between Alice and Bob is $\vert t^{/}_{1}\rangle_{AB}$. Alice introduces auxilliary qubit $\vert 0\rangle_{aux}$ and takes into consideration the unitary operator (\ref{unitarymatrix3}) such that the collective unitary operation $U^{/}_{1} \otimes I_{B}$ transforms the state $\vert t^{/}_{1}\rangle_{AB} \otimes \vert 0\rangle_{aux}$ to $\vert t^{/}_{1}\rangle_{ABaux}$ as before, and subsequently if we assume that Alice's von-Neumann measurement outcome is $\vert 0\rangle_{aux}$, then shared state between Alice and Bob is given by\\
\begin{eqnarray}
\sin\:\theta\:\cos\:\varepsilon\:\vert 01\rangle_{AB} +\cos\:\theta\:\cos\:\varepsilon\:\vert 10\rangle_{AB}\nonumber{}\\
+ (\sin\:\theta\:\sin\:\varepsilon\: + \frac{\sin^{2}\theta\:\cos\:\varepsilon}{\cos\:\theta})\vert 00\rangle_{AB}.
\label{W4}
\end{eqnarray}\\\\\\
Using (\ref{concur1}), we calculate the concurrence ($C_{3}$) of the state (\ref{W4}) and when we plot this concurrence against $\theta$ and $\varepsilon$, we find the following picture.
\begin{figure}[!ht]
\centering
\resizebox{7cm}{7cm}{\includegraphics{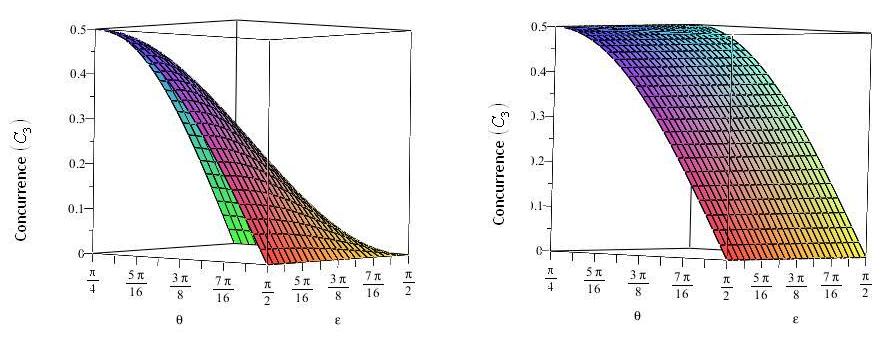}}
\caption{\footnotesize Here $\frac{\pi}{4} \leq \theta \leq \frac{\pi}{2}$ and $\frac{\pi}{4} \leq \varepsilon \leq \frac{\pi}{2}$, whereas $0 \leq C_{3} \leq 0.5$}
\end{figure}\newpage
In Fig-$6$, the first plot represents the concurrence $C_{3}$ against $\:\lbrace \varepsilon, \theta\:\rbrace$. The second plot of Fig-$6$ gives the variation of $C_{3}$ against $\theta$ whereas $\varepsilon = \frac{\pi}{4}$.  From the above analysis we see that, the state (\ref{W1}) is also not suitable  for controlled dense coding as because when Paul and Charlie varies their parameter $\varepsilon$ and $\theta$, the concurrence of the state shared between Alice and Bob reaches its maximum value upto $0.5$ and therefore the shared state is never maximally entangled. Also if we consider the shared state between Alice and Bob corresponding to the states (\ref{w1}) and (\ref{W1}), we find that although the entanglement of the shared state of Alice and Bob for the state (\ref{w1}) is more than that for the state (\ref{W1}),it is also not useful for controlled dense coding.
\subsection{$W_{n}$ state:}
We in this section now consider the state 
\begin{equation}
\vert W_{n}\rangle_{ABC}= \frac{1}{\sqrt{2}}[\vert \phi\rangle_{AB}\:\vert 0\rangle_{C}+\vert 00\rangle_{AB}\:\vert 1\rangle_{C}]\label{W1n}.
\end{equation}
where
\begin{equation}
\vert \phi\rangle_{AB}= \frac{1}{\sqrt{n+1}}[\vert 10\rangle_{AB}+\sqrt{n}\vert 01\rangle_{AB}]\label{W2n}.
\end{equation}
This state can be converted from  $\vert GHZ\rangle_{ABC}=\frac{\vert 000\rangle_{ABC} + \vert 111\rangle_{ABC}}{\sqrt{2}}$ \cite{liqiu} as shown below
\begin{equation}
\vert W_{n}\rangle_{ABC}= (U_{AB} \otimes I_{C})\,\vert GHZ\rangle_{ABC}\label{convert1}
\end{equation}
while $U_{AB}$ is a unitary operator acting on particles A and B given as 
\begin{equation}
U_{AB}= \vert \phi\rangle\langle 00\vert + \vert 11\rangle\langle 01\vert + \vert \phi^{\bot}\rangle\langle 10\vert + \vert 00\rangle\langle 11\vert \label{convert2}.
\end{equation}
Phase factors have not been considered here.
When Charlie considers his basis (\ref{charliebasis1}), $\vert W_{n}\rangle_{ABC}$ takes the following form
\begin{eqnarray}
\vert W_{n}\rangle_{ABC}= \frac{1}{\sqrt{2}}[(\cos\:\theta\:\vert \phi\rangle_{AB}+\sin\:\theta\:\vert 00\rangle_{AB})\:\vert +\rangle_{C}\:\nonumber{}\\
+ (\sin\:\theta\:\vert \phi\rangle_{AB}-\cos\:\theta\:\vert 00\rangle_{AB})\:\vert -\rangle_{C}]\label{W3n}
\end{eqnarray}
If Charlie's measurement outcome is $\vert +\rangle_{C}$, then the state shared between Alice and Bob is
\begin{equation}
\vert \upsilon\rangle_{AB} = \cos\:\theta\:\vert \phi\rangle_{AB} + \sin\:\theta\:\vert 00\rangle_{AB} \label{W4n}
\end{equation}
Alice introduces auxilliary qubit $\vert 0\rangle_{aux}$ and the collective unitary operation $U_{3} \otimes I_{B}$ transforms $\vert \upsilon\rangle_{AB} \otimes \vert 0\rangle_{aux}$ in to the following state
\begin{eqnarray}
\vert \upsilon\rangle_{ABaux} = [\frac{\sqrt{n}}{\sqrt{n+1}}\:\cos\:\theta\:\vert 01\rangle_{AB}+\sin\:\theta\:\vert 00\rangle_{AB}+\frac{\sin\:\theta}{n+1}\vert 10\rangle_{AB}] \otimes \vert 0\rangle_{aux} \nonumber{}\\
-\frac{1}{\sqrt{n+1}}\cos\:\theta\:\sqrt{1-\frac{\sin^{2}\theta}{\cos^{2}\theta}}\vert 00\rangle_{AB}\otimes \vert 1\rangle_{aux}\label{W5n}
\end{eqnarray}
where we take the form of $U_{3}$ as
\begin{eqnarray}
U_{3} = \left(%
\begin{array}{cccc}
1& 0& 0& 0\\
0& \frac{\sin\:\theta}{\cos\:\theta}& \sqrt{1-\frac{\sin^{2}\:\theta}{\cos^{2}\:\theta}}& 0\\
0& -\sqrt{1-\frac{\sin^{2}\:\theta}{\cos^{2}\:\theta}}& \frac{\sin\:\theta}{\cos\:\theta}& 0\\
0& 0& 0& 1
\end{array}%
\right)\label{unitarymatrix5}
\end{eqnarray}
For Alice's von-Neumann measurement outcome $\vert 0\rangle_{aux}$, we see that the shared state between Alice and Bob takes the form
\begin{equation}
\frac{\sqrt{n}\:\cos\:\theta\:\vert 01\rangle_{AB}+\sin\:\theta\:\vert 10\rangle_{AB}}{\sqrt{n+1}} + \sin\:\theta\:\vert 00\rangle_{AB}\label{W6n}
\end{equation}
For $n=1$, (\ref{W6n}), takes the form
\begin{equation}
\frac{\cos\:\theta\:\vert 01\rangle_{AB}+\sin\:\theta\:\vert 10\rangle_{AB}}{\sqrt{2}} + \frac{\sqrt{2}\:\sin\:\theta\:\vert 00\rangle_{AB}}{\sqrt{2}}\label{W7n}
\end{equation}
Hence Alice and Bob shares non-maximally entangled state $\cos\:\theta\:\vert 01\rangle_{AB}+\sin\:\theta\:\vert 10\rangle_{AB}$ with probability $\frac{1}{2}$. Using (\ref{concur1}) we calculate the concurrence $C_{4}$ of the state $\cos\:\theta\:\vert 01\rangle_{AB}+\sin\:\theta\:\vert 10\rangle_{AB}$. \\\\
\begin{equation}
C_{4}=\vert\sin\:2\:\theta\vert.
\end{equation}\\\\
If we plot $C_{4}$ against $\theta$, we get the following\\
\begin{figure}[!ht]
\centering
\resizebox{7cm}{7cm}{\includegraphics{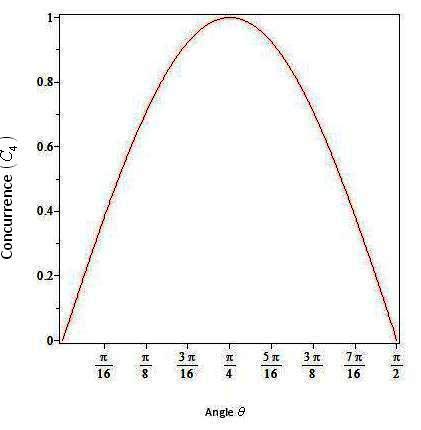}}
\caption{\footnotesize Here $0 \leq \theta \leq \frac{\pi}{2}$, $0 \leq C_{4} \leq 1$ and $C_{4}$ takes its maximum value $1$ for $\theta=\frac{\pi}{4}$}
\end{figure}\newpage
From the above it is clear that controlled dense coding with states of the form (\ref{W1n}) is possible when Charlie's measurement angle takes the value $\frac{\pi}{4}$ as in that case Alice and Bob shares maximally entangled state $\frac{\vert 01\rangle_{AB}+\vert 10\rangle_{AB}}{\sqrt{2}}$.

\section{Controlled dense coding with tripartite qutrit state:}
Here we consider a tripartite qutrit state from \cite{liulongtongli} of the following form
\begin{equation}
\vert \Xi\rangle_{ABC} = \frac{\vert 000\rangle_{ABC} +\vert 111\rangle_{ABC} + \vert 222\rangle_{ABC}}{\sqrt{3}}\label{qutritstate1}
\end{equation}
A, B, C are Alice, Bob and Charlie each possessing three qubits $\vert 0\rangle$, $\vert 1\rangle$ and $\vert 2\rangle$ respectively. This state is basically a natural generalization of GHZ state in three level systems with full rank of all reduced density matrices \cite{florian}. Charlie now considers the following basis
\begin{eqnarray}
\vert \uparrow\rangle _{C} &=& \sin\:\theta\vert 0\rangle_{C} + \cos\:\theta\vert 2\rangle_{C},\nonumber{}\\
\vert \nearrow\rangle_{C}&=&\vert 1\rangle_{C},\nonumber{}\\
\vert \downarrow\rangle_{C} &=& \cos\:\theta\vert 0\rangle_{C} - \sin\:\theta\vert 2\rangle_{C}. \label{charliebasis2}
\end{eqnarray}
With respect to Charlie's basis (\ref{charliebasis2}), the state (\ref{qutritstate1}) takes the form
\begin{eqnarray}
\vert \Xi\rangle_{ABC} = (\sin\:\theta\vert 00\rangle_{AB}+\cos\:\theta\vert 22\rangle_{AB})\vert \uparrow\rangle_{C}\nonumber{}\\
+(\cos\:\theta\vert 00\rangle_{AB}-\sin\:\theta\vert 22\rangle_{AB})\vert \downarrow\rangle_{C}+ \vert 11\rangle_{AB}\vert \nearrow\rangle_{C}.\label{qutritstate2}
\end{eqnarray}\\
When Charile obtains measurement result $\vert \nearrow\rangle_{C}$, then Alice and Bob shares $\vert 11\rangle_{AB}$ and only one bit is transferred from Alice to Bob in that case. But when charlie's measurement basis is $\lbrace\:\vert \uparrow\rangle_{C}, \vert \downarrow\rangle_{C}\:\rbrace$, then  a non-maximally entangled shared state is shared between them. Let Charlie's measurement result is $\vert \uparrow\rangle_{C}$, so that the shared entangled state between Alice and Bob is ($\sin\:\theta\vert 00\rangle_{AB}+\cos\:\theta\vert 22\rangle_{AB}$).\\\\
Alice introduces an auxiliary qubit $\vert 0\rangle_{aux}$ and performs an unitary operation on her qubit A as well as on the auxiliary qubit (with respect to the collective operation under the basis ($\vert 00\rangle_{Aaux}, \vert 01\rangle_{Aaux}, \vert 02\rangle_{Aaux}, \vert 10\rangle_{Aaux},\\ \vert 11\rangle_{Aaux}, \vert 12\rangle_{Aaux}, \vert 20\rangle_{Aaux}, \vert 21\rangle_{Aaux}, \vert 22\rangle_{Aaux}$ ). In this case we consider we consider the $9 \times 9$ unitary Braid matrix \cite{abdchakdobmihov}.\\
\begin{eqnarray}
B_{1} = \left(%
\begin{array}{ccccccccc}
\frac{\cos\:\theta}{\sin\:\theta}& 0& 0& 0& 0& 0& 0& 0& \sqrt{1-\frac{\cos^{2}\:\theta}{\sin^{2}\:\theta}}\\
0& 1& 0& 0& 0& 0& 0& 0& 0\\
0& 0& \frac{\sin\:\theta}{\cos\:\theta}& 0& 0& 0& \sqrt{1-\frac{\sin^{2}\:\theta}{\cos^{2}\:\theta}}& 0& 0\\
0& 0& 0& 1& 0& 0& 0& 0& 0\\
0& 0& 0& 0& 1& 0& 0& 0& 0\\
0& 0& 0& 0& 0& 1& 0& 0& 0\\
0& 0& \sqrt{1-\frac{\sin^{2}\:\theta}{\cos^{2}\:\theta}}& 0& 0& 0& -\frac{\sin\:\theta}{\cos\:\theta}& 0& 0\\
0& 0& 0& 0& 0& 0& 0& 1& 0\\
\sqrt{1-\frac{\cos^{2}\:\theta}{\sin^{2}\:\theta}}& 0& 0& 0& 0& 0& 0& 0& -\frac{\cos\:\theta}{\sin\:\theta}
\end{array}%
\right).\label{unitarymatrix6}
\end{eqnarray}\\\\
Then the collective unitary operation $B_{1} \otimes I_{B}$ transforms the state $(\sin\:\theta\vert 00\rangle_{AB}+\cos\:\theta\vert 22\rangle_{AB}) \otimes \vert 0\rangle_{aux}$ to the state
\begin{eqnarray}
\vert \Phi\rangle_{ABaux}=(\cos\:\theta\vert 00\rangle_{AB}-\sin\:\theta\vert 22\rangle_{AB})\otimes \vert 0\rangle_{aux}\nonumber{}\\
+\lbrace\: \cos\:\theta\:\sqrt{1-\frac{\sin^{2}\theta}{\cos^{2}\theta}}\vert 02\rangle_{AB} + \sin\:\theta\:\sqrt{1-\frac{\sin^{2}\theta}{\cos^{2}\theta}}\vert 20\rangle_{AB}\:\rbrace \otimes \vert 2\rangle_{aux}. \label{qutritstate3}
\end{eqnarray}
Now for Alice's von - Neumann readout $\vert 0 \rangle_{aux}$, the non-maximally entangled state shared between Alice and Bob is ~~$\cos\:\theta\vert 00\rangle_{AB} - \sin\:\theta\vert 22\rangle_{AB}$. It immediately follows that when $\theta=\frac{\pi}{4}$, then Alice and Bob share the maximally entangled state
\begin{equation}
\frac{\vert 00\rangle_{AB}-\vert 22\rangle_{AB}}{\sqrt{2}}.\label{qutritstate4}
\end{equation}
Alice applies the projection operators $\vert 0\rangle\langle 0\vert + \vert 2\rangle\langle 2\vert$, $\vert 0\rangle\langle 2\vert + \vert 2\rangle\langle 0\vert$, $\vert 0\rangle\langle 2\vert - \vert 2\rangle\langle 0\vert$ and $\vert 0\rangle\langle 0\vert - \vert 2\rangle\langle 2\vert$ respectively to the state (\ref{qutritstate4}), and she obtains the following states.
\begin{eqnarray}
\frac{\vert 00\rangle_{AB}-\vert 22\rangle_{AB}}{\sqrt{2}},\nonumber{}\\
\frac{\vert 20\rangle_{AB}-\vert 02\rangle_{AB}}{\sqrt{2}},\nonumber{}\\
\frac{\vert 02\rangle_{AB}+\vert 20\rangle_{AB}}{\sqrt{2}},\nonumber{}\\
\frac{\vert 00\rangle_{AB}+\vert 22\rangle_{AB}}{\sqrt{2}}.\label{qutritstate5}
\end{eqnarray}
After this she sends her qubit to Bob. Bob then uses a projection operator $\vert 00\rangle\langle 00\vert + \vert 22\rangle\langle 20\vert + \vert 02\rangle\langle 02\vert + \vert 20\rangle\langle 22\vert$, to his qubit. Bob's projection operator can also be shown in a matrix form in the following
\begin{eqnarray}
U_{P} = \left(%
\begin{array}{ccccccccc}
1& 0& 0& 0& 0& 0& 0& 0& 0\\
0& 0& 0& 0& 0& 0& 0& 0& 0\\
0& 0& 1& 0& 0& 0& 0& 0& 0\\
0& 0& 0& 0& 0& 0& 0& 0& 0\\
0& 0& 0& 0& 0& 0& 0& 0& 0\\
0& 0& 0& 0& 0& 0& 0& 0& 0\\
0& 0& 0& 0& 0& 0& 0& 0& 1\\
0& 0& 0& 0& 0& 0& 0& 0& 0\\
0& 0& 0& 0& 0& 0& 1& 0& 0
\end{array}%
\right)\label{unitarymatrix7}
\end{eqnarray}\\\\
By measuring qubit $A$ under the basis $\lbrace\:\vert 0\rangle_{A} \pm \vert 2\rangle_{A}\:\rbrace$ and $\lbrace\:\vert 2\rangle_{A} \pm \vert 0\rangle_{A}\:\rbrace$, he can distinguish Alice's operations and the phase bit is gotten, so $2$ bits, $\lbrace\:0, 2\:\rbrace$, of information are transmitted. If, however the shared state between Alice and Bob is $(\cos\:\theta\vert 00\rangle_{AB}-\sin\:\theta\vert 22\rangle_{AB})$, then Alice will consider the following form of the Braid matrix \cite{abdchakdobmihov}
\begin{eqnarray}
B_{2} = \left(%
\begin{array}{ccccccccc}
\frac{\sin\:\theta}{\cos\:\theta}& 0& 0& 0& 0& 0& 0& 0& \sqrt{1-\frac{\sin^{2}\:\theta}{\cos^{2}\:\theta}}\\
0& 1& 0& 0& 0& 0& 0& 0& 0\\
0& 0& \frac{\cos\:\theta}{\sin\:\theta}& 0& 0& 0& \sqrt{1-\frac{\cos^{2}\:\theta}{\sin^{2}\:\theta}}& 0& 0\\
0& 0& 0& 1& 0& 0& 0& 0& 0\\
0& 0& 0& 0& 1& 0& 0& 0& 0\\
0& 0& 0& 0& 0& 1& 0& 0& 0\\
0& 0& \sqrt{1-\frac{\cos^{2}\:\theta}{\sin^{2}\:\theta}}& 0& 0& 0& -\frac{\cos\:\theta}{\sin\:\theta}& 0& 0\\
0& 0& 0& 0& 0& 0& 0& 1& 0\\
\sqrt{1-\frac{\sin^{2}\:\theta}{\cos^{2}\:\theta}}& 0& 0& 0& 0& 0& 0& 0& -\frac{\sin\:\theta}{\cos\:\theta}
\end{array}%
\right)\label{unitarymatrix8}
\end{eqnarray}\\\\
Applying the protocol as described above, $2-$ bits of information are transferred from Alice to Bob again.
\section{Summary and Discussion:}
To summarize, we have discussed the scheme of controlled dense coding with different types of tripartite and quadrpartite states. The GHZ states and its various similar classifications have been considered first, in which case the number of bits transmitted is $2$. It is found that, the average number bits transmitted from sender to receiver depends on the basis of measurements chosen by the controller of the scheme. It has been shown that GHZ - type states like $L(\vert 000\rangle + l\:\vert 111\rangle)$, where $L\:=\frac{1}{\sqrt{1 + l^{2}}}$ can be used in controlled dense coding scheme in a probabilistic sense. We have also found that for these GHZ type states , Charlie (the controller of the scheme) chooses parameter $l$ by manipulating his measurement angle $\theta$. Consequently we have found a relation between $l$ and $\theta$. Also a relation between concurrence of the bipartite state shared by Alice-Bob with respect to the state $L(\vert 000\rangle + l\:\vert 111\rangle)$ and measurement angle $\theta$ of Charlie have been established. It is shown that for $\theta = \frac{\pi}{4}$, the controlled dense coding scheme is achieved successfully with success probability $1$. Controlled dense coding is also achieved with Maximal Slice state only if the parameters $\theta, ~~\delta$ are chosen in an appropriate way. For proper chioces of $\theta, ~~\delta$, the maximal sliced states basically takes the GHZ form. Next four particle GHZ - state is also taken into consideration and the relation between concurrence of the shared bipartite state by Alice-Bob  and parameters $\lbrace\:\theta, \varepsilon\:\rbrace$ is established. The study of controlled dense coding is made with W-class states. By considering the bipartite state shared by Alice - Bob for the W-class of states and by calculating the concurrence of these states, we have explained that states like $\vert W \rangle_{ABC}$ and $\vert W \rangle_{PABC}$ cannot be used in controlled dense coding. It is because, the maximum value of concurrence for the shared bipartite states  with respect to $\vert W \rangle_{ABC}$   and  $\vert W \rangle_{PABC}$ are $0.7$ and $0.5$ respectively and never reaches its maximal value $1$. But it is interesting to see that the state $\vert W_{n} \rangle_{ABC}$ can be used successfully in controlled dense coding in a probabilistic manner for parameter values $n=1$ and $\theta=\frac{\pi}{4}$. Finally, we have successfully performed the controlled dense coding scheme on tripartite qutrit state of GHZ form, by introducing two $9 \times 9$ unitary Braid matrices to send $2$ classical bits of information from Alice to Bob controlled by third party Charlie. So we can infer that the tripartite qutrit state of GHZ form is an optimal state that achieve the desired task. In future it would be quite interesting to make a quantitative study of this above scheme with GHZ states in arbitrary dimension and for maximally and non - maximally entangled mixed states which is very important from quantum information theoretic perspective. 
\section*{Acknowledgement:}
 The authors Roy and Ghosh Acknowledge Dr. Satyabrata Adhikari (associate professor in the department of Mathematics) of Birla Institute of Technology, Mesra and Dr. Archan S. Majumdar (associate professor in the department of Astrophysics and Cosmology) of S. N. Bose National Centre of Basic Sciences, who were the co-authors of our earlier work on MEMS and NMEMS states, for their valuable inputs, time to time suggestions and consistent support.

\end{document}